\documentclass[useAMS,usenatbib,consecutivenumbering,epsf,psfig]{mn2e}
\usepackage{amsmath, amssymb, amsfonts, amssymb,epsfig}
\usepackage{url}

\begin{document}
\setcounter{table}{0}
\setcounter{figure}{0}
\def\msun{{\rm M}_\odot}
\def\Msun{{\rm M}_\odot}
\def\rsun{{\rm R}_\odot}
\newcommand{\lsimeq}{\mbox{$\, \stackrel{\scriptstyle <}{\scriptstyle
\sim}\,$}}
\newcommand{\gsimeq}{\mbox{$\, \stackrel{\scriptstyle >}{\scriptstyle
\sim}\,$}}
\def\gradi{\ifmmode{^\circ}\else$^\circ$\fi}

\title[White dwarfs in binaries]{Constraints on the pairing properties
  of main sequence stars from observations of white dwarfs in binary
  systems.}

\author[L. Ferrario]
{Lilia Ferrario \\
Mathematical Sciences Institute, The Australian National University,
Canberra, ACT 0200, Australia}

\date{Accepted. Received ; in original form}

\maketitle
\begin{abstract}
Observations of main sequence stars conducted over the last several
decades have clearly shown that something like 50 per cent of stars of
spectral types G and F occur in multiple systems. For earlier spectral
types, the incidence of multiplicity is even higher. Thus, a volume
limited sample of white dwarfs should reflect the percentage of
binarity observed in stars of F to late B spectral types, which are
their Main Sequence progenitors. However, a study of the local volume
limited sample of white dwarfs (20 pc from the Sun) conducted by
Holberg has shown that a white dwarf has a probability of only $\sim
32$ per cent of occurring in a binary system, in stark contrast to the
observations of multiplicity of Main Sequence stars. Others studies
have also led to the same conclusion.

In this paper, we argue that the ``hidden'' white dwarfs are either in
double white dwarf systems or in Sirius-like systems. We also show
that the white dwarf progenitors of the SDSS white dwarf - M dwarf
wide binaries are distributed according to Salpeter's IMF. However,
they cannot be paired with secondary stars which are also drawn from
this IMF, since such a pairing would produce a percentage of white
dwarf-M dwarfs systems that is several times larger than observed.

\end{abstract}

\begin{keywords}
White dwarfs, Binaries: general, stars: formation, stars: low-mass.
\end{keywords}

\section{Introduction}

White dwarf-M dwarf pairs are commonly discovered in any surveys of
white dwarfs. The observed pairs fall into three basic groups. In the
first group (group A), the more massive star evolves into a white
dwarf without ever interacting with its companion. These are the
so-called wide binaries.  In the second group (group B), the binary
stars have smaller initial separations and will eventually evolve
through a common envelope (CE) phase as the more massive star becomes
a white dwarf. These are the post common envelope pre-Cataclysmic
Variable binaries (Pre CVs). The third group (group C) consists of
pairs that have already evolved through a CE phase and are currently
seen as close interacting binary systems where there is evidence of
mass transfer (current or in the past). Thus, the M dwarfs in these
systems are either filling or close to filling their Roche
lobes. These are the Cataclysmic Variable (CVs) stars.

Studies aimed at determining the observed distribution $f_{\rm
  obs}(q)$ of mass ratios $q=\displaystyle{\frac{M_s}{M_p}}<1$, where
$M_p$ is the mass of the primary star and $M_s$ is the mass of its
less massive companion, are generally limited to establishing the
nature of the distribution for $q\ge 0.1$ when the luminosity of the
secondary star is not swamped by that of the primary.  On the other
hand, insights into the behaviour of $f_{\rm obs}(q)$ at low values of $q$ can
be more easily obtained from studies of the pairing of white dwarfs
with main sequence companions belonging to group A and through simple
assumptions on the initial-final mass relationship for white dwarfs.

In this paper, we analyse the pairing properties of white dwarf - Main
Sequence binaries using as constraints the white dwarfs - M dwarf
sample from the SDSS belonging to group A (wide binaries), the
\citet{Holberg09} percentage of observed white dwarf binaries in the
20 pc local sample and observations of the mass ratio distribution
of binary Main Sequence stars. We then present some conclusions on the
pairing properties of Main Sequence low to intermediate mass stars,
which are the progenitors of the currently observed white dwarfs, and
link them to possible star formation scenarios.

\section{The observational basis}

\subsection{Mass ratio distribution of binary main sequence stars}\label{mass_ratio_MS}

Early studies of binary systems pointed to a bi-modal distribution of
$f_{\rm obs}(q)$ (e.g. \citet{Trimble74}): a population $f_{\rm
  obs}(q)$ which increases with increasing $q$ and reaches a maximum
near $q=1$ thus showing a preference for stars of similar masses
(twins), and a second population $f_{\rm obs}(q)$ which initially
rises towards lower $q$, reaches a maximum at $q\sim 0.2$ and then
decreases.  More recent studies using F7 to K-type primary stars by
\citet{Halbwachs03} have shown that $f_{\rm obs}(q)$ has a very broad peak at
$q\sim 0.2-0.7$ (perhaps with sub-structure) and a sharp peak for
$q\gsimeq 0.8$ (twins). Interestingly, these authors claim that the
$f(q)$ relationship is ``scale-free'' for stars in the three different
sub-groups that comprised their F7-K star sample.

\citet{Kiminki12} conducted a statistical analysis of massive binaries
in the Cygnus OB2 association using radial velocity data for 114
B3--O5 primary stars. They found a mass ratio distribution
$f_{\rm obs}(q)\propto q^\alpha$, with $\alpha= 0.1\pm 0.5$, which is
consistent with their previous work \citep{Kobulnicky07}. This is also
in agreement with the observations of \citet{Kouwenhoven05}, who find
that A and late B-type stars in the Scorpius OB2 association have a
mass ratio distribution with $\alpha=-0.33$, and with the studies of
\citet{Shatsky02}, who also surveyed B-type stars in Scorpius OB2 for
binarity and found $\alpha$ in the range $-0.3$ to $-0.5$.  Studies of
a sample of massive binaries in the Small Magellanic Cloud conducted
by \citet{Pinsonneault06} revealed that their primaries appear to have
two populations of companions: a ``twin'' population with $q>0.95$
comprising 45 per cent of binaries, and a population with
$f_{\rm obs}(q)\sim {\mbox{constant}}$ (i.e. $\alpha=0$) comprising 55 per
cent for their sample. On the other hand, \citet{Sana11} claim that
there is no indication for a twin population in their Galactic O-star
sample and that the mass-ratio distribution is essentially flat for
$0.2<q<1.0$.  The studies by \citet{Raghavan12} on companions to
solar-type stars indicate that the mass ratio distribution is flat for
$\sim 0.2\le q\le 0.95$.

\citet{Woitas01} studied the masses and mass ratios of pre-Main
Sequence stars. Interestingly, they also found that the distribution
of mass ratios in T Tauri binaries is essentially flat for $q>0.2$ and
it has no correlation with the primary's mass or the stellar
separation. They also found that there is no significant preference
for $q\sim 1$ (twin component). 

In summary, at the present time it appears that the $f(q)$
distribution is either flat or slightly raising toward $q\sim 0.1-0.3$
with or without a rise toward $q=1$ (twin). Interestingly, these
studies appear to agree for primaries ranging from solar to early
O-type stars and even for young protostellar associations.  However,
none of these studies extends to low enough mass ratios to investigate
the true incidence of binaries whose secondaries are M dwarfs.

\subsection{Observations of white dwarf - main sequence star binaries}\label{WDMS}

While an M-dwarf is easily hidden in the glare of an intermediate or
high-mass star primary, such a low luminosity star should be more
easily seen after its more massive companion has evolved into a
compact star. Thus, insights into the behaviour of $f(q)$ at low $q$
can potentially be obtained from studies of the pairing of white
dwarfs with low-mass main sequence stars and by assuming an initial to
final mass relationship for white dwarfs (e.g. \citet{Ferrario05}).

A rather surprising result resides in the observed incidence of
binarity among white dwarfs. \citet{Farihi05} found that the stellar
companion fraction of white dwarfs is 22 per cent, uncorrected for
bias. Furthermore, they also found that most of the stellar companions
to white dwarfs are low-mass M dwarf stars.  Similarly,
\citet{Holberg09} found that only $32\pm 8$ per cent of the local
white dwarf population has a companion of \emph{any} type.

However, it is well known that the progenitors of the currently
observed white dwarfs, which are Main Sequence stars in the mass range
$\sim 1.2-8 \Msun$, exhibit a percentage of binarity of at
least 55 per cent at the low mass limit (F type stars), to 60 per cent
or more toward the upper end of the mass range limit (late B type
stars). Thus, there must be an additional $\sim 30$ per cent of as yet
undiscovered white dwarfs lurking in some kind of binaries.

Interestingly, if one assumes that the majority of white dwarf-M dwarf
binaries in the local sample has already been detected, since any M
dwarf red excess would be clearly visible in the spectrum of a nearby
white dwarf, then according to the percentages given by
\citet{Holberg09}, the incidence of this type of pairing could be
\emph{as low as} $\sim 18$ per cent, consistent with the findings of
\citet{Farihi05}. Considering that M-dwarfs are the most numerous
stars in the Galaxy, the fact that they are somewhat rarely paired to
white dwarfs, suggests that binary formation mechanisms tend to
exclude them as companions of F to late B-type stars.

Another possibility that cannot be excluded a priori is that the local
white dwarf binary population is not representative of the true
Galactic population. If this is the case, our calculations, which
use as constraint current observations of Galactic stellar
multiplicity, will simply give predictions that would be applicable to
a complete white dwarf binary sample.

\begin{figure}
\begin{center}
\hspace{0.1in}
\epsfxsize=0.98\columnwidth
\epsfbox{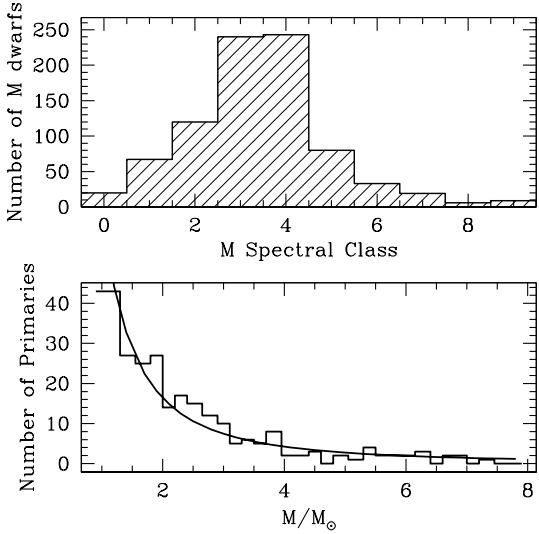}
\caption{Top panel: Spectral class of M dwarf companions to white
  dwarfs in the wide binaries of \citet{Rebassa10}. Bottom panel: Mass
  distribution of white dwarf progenitors after a final-to-initial
  mass relationship has been applied to the white dwarfs in the wide
  binaries in \citet{Rebassa10} (see text for details). Solid curve:
  best power law fit with index $\alpha=-1.95\pm 0.14$.}
\label{rebassa_data}
\end{center}
\end{figure}

We show in Fig. (\ref{rebassa_data}) the spectral distribution of M
dwarfs found in wide binaries containing a white dwarf from the
spectroscopic SDSS Data Release 6 \citep{Rebassa10}.

We note that the peak in the M dwarf secondary distribution is near
M3.5 and that there is a very steep decline in the number of M dwarfs
of spectral type later than M5. Interestingly, \citet{Farihi05} also
find that the peak frequency in spectral type occurs around M3.5 for
\emph{both} field M dwarfs and M dwarf companions to white dwarfs (see
their Figures 6 and 7). However, relatively to the peak, they find
that there are $\sim 2-3$ times more L dwarfs and $\sim 4-5$ times
more M6–M9 dwarfs in the field than among companions. Thus, despite
the excellent sensitivity in their survey to late-M dwarfs and early-L
dwarfs very few of these low-mass companions were detected.

We note that the SDSS white dwarf - M dwarf binary data suffer from
strong selection effects, since the SDSS is a magnitude-limited
survey. Thus, one should expect that selection effects may become dominant at
later spectral types (M5 to M9) if the white dwarf is more luminous
than the M dwarf, or at early spectral type (M0 to M2), when the fainter
white dwarfs may be hidden by the more luminous M dwarfs.  Therefore, one
may expect that even after introducing corrections for observational
biases, the peak of the distribution would remain around M3.5. We
shall use this peak as a constraint for our studies in section
(\ref{results}).

An important piece of information that can be extracted from the data
of \citet{Rebassa10} concerns the distribution of white dwarf masses -
and thus of their Main Sequence progenitor masses. Since we need an
initial-to-final mass function to obtain the masses of the WD
progenitors, we have used the simple relation of
\citet{Catalan09}. The histogram of the mass distribution that we have
obtained is shown in the bottom panel of Fig.
(\ref{rebassa_data}). This histogram shows that the number of WD
progenitors drops very quickly with mass. We have found that the best
power law fit to these data has an index $\alpha=-1.95\pm 0.14$,
consistent with the observational results of \citet{Chabrier03} who
found a slope $\alpha=-2.3\pm 0.3$ for single stars with $M>1\Msun$ in
the galactic disk and young clusters.

In the next sections, we investigate pairing functions of main
sequence stars. We will show that given the current observational
constraints, there should be a sizeable fraction of WDs in Sirius-type
systems and in double white dwarf binaries.

\section{Calculations}\label{calculations}

In order to investigate the properties of white dwarfs in binaries, a
number of systems was generated with primary masses distributed
according to 
\[
\frac{dN}{d\log M_p} \propto M_p^{-0.95},
\]
as indicated by the SDSS data (see previous section). The
justification for this is that the distribution of primary masses
should exactly reflect the distribution of their masses at birth,
since in these wide binaries the two stellar components have never
come into contact and exchanged mass. One should also keep in mind
that the mass range of the primaries is automatically restricted to
$1.2 \lsimeq M_p\lsimeq 8 M_\odot$ since the Galactic disk is not old
enough for lower mass stars to have evolved to the white dwarf stage,
while stars with $M\gsimeq 8 \msun$ undergo supernova
explosions and become neutron stars.

In our calculations, we have assumed that either (i) the masses of the
two stars are both drawn from a mass distribution $f(M_s)$ with the
constraint $M_s\le M_p$ ($q\le 1$) or (ii) the mass of the secondary
star depends on the mass of the primary, or (iii) the binaries are
generated according to a Salpeter's like distribution for the primary
mass and with the secondary mass determined by a generating mass ratio
distribution $f(q)$. Thus we have explored the following pairing cases

\begin{itemize}

\item [(i)]
A distribution for the mass of the secondaries given by
\begin{equation}\label{powerlaw}
\frac{dN}{d\log M_s}\propto M_s^\rho
\end{equation}
where $dN$ is the number of stars per unit volume per logarithmic mass
interval $d\log M_s$ observed at time $t$. If $\rho=-1.35$, we have the
\citet{Salpeter55} mass function. 

\item [(ii)] A distribution for the mass of the secondaries given by
\begin{equation}\label{bochanski}
\frac{dN}{d\log M_s}\propto
\begin{cases}
\exp\left[\displaystyle{\frac{-(\log M_s-\log M_c)}{2\sigma^2}}\right] & \displaystyle{\frac{M_s}{\Msun}}\le 1, \\
M_s^\delta &  1<\displaystyle{\frac{M_s}{\Msun}}<8.
\end{cases}
\end{equation}
This type of distribution was first suggested by \citet{Miller79} and was
more recently adopted by, e.g., \citet{Chabrier05} and
\citet{Bochanski10}. Here, $\log M_c$ and $\sigma$ are the average
mass and standard deviation, respectively, in $\log M_s$. In the
distribution of \citet{Bochanski10} (used in this paper), $\log
M_c=-0.745$ and $\sigma=0.34$.

\item [(iii)] A Gaussian distribution where the mass of the secondary is
  proportional to the mass of the primary star $M_p$:

\begin{equation}\label{gauss}
\frac{dN}{d\log M_s}\propto \exp\left[\frac{-(\log M_s-\log (kM_p))}{2\sigma^2}\right]
\end{equation}
where $k$ is the proportionality constant.

\item [(iv)] The mass of the secondary is determined by the mass ratio that is
  drawn from a generating mass ratio distribution $f(q)$ given
  by
\begin{equation}\label{qratio_1}
f(q) \propto q^\beta \qquad 0< q\le 1 \\
\end{equation}

\end{itemize}
We note that if $\tau_D$ is the age of the galactic disk, assumed to
be 9.5 Gyr (e.g. \citet{Oswalt96}), and $\tau_{\rm prewd}(M_i)$ is the
nuclear burning evolutionary time of a star of initial mass $M_i$,
then only those stars born at a time $0\le \tau \le \tau_D-\tau_{\rm
  prewd}$ can currently be observed as white dwarfs of age $\tau_{\rm
  wd}=\tau_D-\tau_{\rm prewd}-\tau$. In our calculations, we have
assumed that the Galactic star formation rate is is constant over the
lifetime of the Galactic disc.  To obtain an estimate of $\tau_{\rm
  prewd}(M_i)$, we have used the average stellar lifetimes given by
\citet{Romano05} based on the stellar evolution grids of
\citet{Maeder89}.

The temperature of the white dwarf can then be calculated from its
age $\tau_{\rm wd}$ and mass $M_{\rm wd}$ by interpolating the tables
for evolutionary sequences of white dwarf atmospheres of
\citet{Holberg06}, \citet{Kowalski06}, \citet{Tremblay11}, and
\citet{Bergeron11}\footnote{\url{http://www.astro.umontreal.ca/~bergeron/CoolingModels}}.

\begin{figure}
\begin{center}
\hspace{0.1in}
\epsfxsize=0.95\columnwidth
\epsfbox[45 150 315 588]{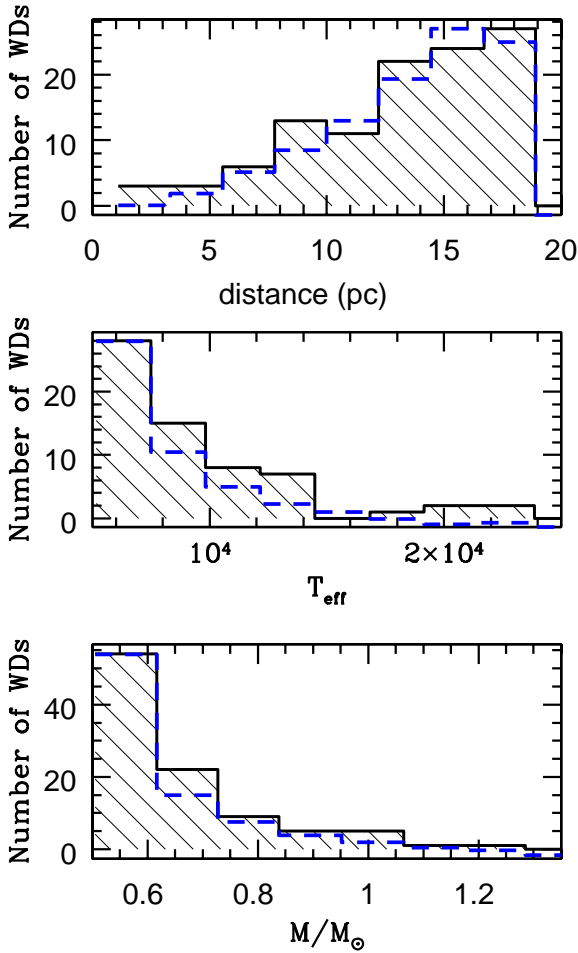}
\caption{From top to bottom: Shaded histograms of
  distance, effective temperature and mass of the 20 pc sample of
  \citet{Holberg09}. Dashed histograms: Synthetic white dwarf data.}
\label{holberg_data}
\end{center}
\end{figure}

Over the evolutionary time of the primary star and subsequent cooling
as a white dwarf, some of the secondaries may also have had enough
time to evolve and to be observed as white dwarfs at the present
epoch. Thus, in addition to systems containing a white dwarf and a Main
Sequence star, there will systems comprising of double white
dwarfs. The mixture of the two will be dictated by the evolutionary
timescales of the secondary star.

The white dwarf - Main Sequence binaries can then be further divided
into systems with low luminosity M dwarf companions and systems with
stars of earlier (K to late B) spectral type companions. The latter
case yields ``Sirius-type'' systems.  If the white dwarf and its main
sequence star form a close system, then the white dwarf could be
hidden in the glare of its main sequence companion (just like Sirius
B). We note that observationally, it is estimated that about 45 per
cent of Sirius-like systems are close or unresolved, with the remainder
being Close Proper Motion systems (Holberg 2012, private
communication).

In order to check whether our calculations are consistent with the
observational characteristics of the local sample of white dwarfs
\citep{Holberg09}, we have plotted in Fig. (\ref{holberg_data}) our
white dwarf synthetic data for distance, effective temperature and
mass overlapped to the 20 pc white dwarf sample of
\citet{Holberg09}. This figure shows that, considering the
approximations outlined above for the stellar models, our synthetic
population of non-interacting white dwarfs is consistent with
observational results of isolated white dwarfs.

\section{Results and discussion}

\subsection{Binary formation mechanisms}\label{form_mech}

Given the complex nature of star formation, it is difficult to predict
a priori the expected distribution of mass ratios in binary systems,
since all stars, including binaries, form in clusters with a multitude
of physical and dynamical factors coming into play.

Over the years, several routes leading to the formation of binaries
have been identified. The main ones being (i) fission, (ii) capture,
(iii) fragmentation. In fission, binaries arise from the slow
contraction of a rotating gas cloud. Under the simplified assumption
of incompressible, non-viscous fluids and of hydrostatic equilibrium,
the contraction of a rapidly rotating object leads to a dumbbell-type
configuration and to the formation of binaries with mass ratios close
to unity. This mechanism was disproved by the numerical calculations
of \citet{Durisen86} who studied the stability of rotating,
\emph{compressible} gas clouds. They showed that the ejection of
matter and the formation of a bar with trailing spiral structures can
efficiently redistribute the angular momentum on a dynamical time
scale avoiding the formation of a binary system.

The breakup of a fast rotating gas cloud while it is collapsing is
referred to as ``prompt fragmentation''. This mechanism was first
proposed by \citet{Hoyle53}. Because of low compressional heating, the
gas can cool radiatively and the contracting gas cloud is initially
isothermal. As the cloud collapses, the density increases and the
Jeans mass decreases causing fragmentation to take place. However, as
contraction continues, the gas cloud's opacity increases causing its
core to heat up. Fragmentation ends once a fragment's temperature, and
thus its Jeans' mass, starts to increase. The critical value of
opacity that halts contraction yields a minimum Jeans' mass value of
about $0.01\msun$. Early computational work on the formation of binary
systems via fragmentation was conducted by \citet{Boss79}. More
recently, \citet{Clarke96} studied the scale-free fragmentation scenario
and its observational implications.

Fragmentation of a protostellar disc, also called ``disc
fragmentation'', occurs once the collapse of the cloud is over and
protostellar objects have already formed. \citet{Bonnell94} proposed
that fragmentation can occur via two mechanisms. One involves
rotational instabilities in a protostellar core. The other invokes the
formation of a rapidly rotating core which is unstable to axisymmetric
perturbations. This core bounces into a ring which quickly fragments
into several components.

The capture mechanism  implies that single stars have already formed and
are still in a (dense) stellar cluster. Binaries are then created when one star
captures another (``dynamical capture'', see \citet{McDonald93}). The
required excess kinetic energy loss could occur via tidal dissipation,
if the stars get sufficiently close to each other, or through
dynamical encounter and transfer of energy to a third star. If capture
occurs early on in the formation process and either one or both young
stars are still surrounded by protostellar discs, then energy
dissipation would effectively occur through discs-stars interaction
(``star-disc capture'', \citet{McDonald95}). 

Clearly, most of the above mechanisms are expected to play a role in a
star forming region, although it is generally expected that cloud
fragmentation may play a dominant role.  This is certainly the picture
that is emerging from the hydrodynamic simulations of star cluster
formation, such as those conducted by \citet{Bate12}, which can
resolve masses down to a few Jupiter masses. Their calculations
produce the \citet{Chabrier05} observed IMF and find that stellar
multiplicity increases with the mass of the primary, in agreement with
observations. Furthermore, they find a mass ratio distribution for
solar-type and M-dwarf binaries which is roughly flat, again
consistent with observations.

It is interesting to note that the flat $f_{\rm obs}(q)$ distribution observed
in protostellar objects seem to support the view that in most multiple
systems in T Tauri associations the components' masses are mainly
determined by the cloud's fragmentation process and not by the
subsequent disc accretion processes \citep{Woitas01}. This finding is
corroborated by the numerical simulations of \citet{Bate97} who have
shown that in the disc accretion stage following cloud fragmentation, the
system's mass ratio would tend to approach unity, which is in
disagreement with the observational work of \citet{Woitas01}. This
result is very significant, since it indicates that binary
characteristics are already determined as early as 1 million years
after cloud fragmentation has begun.

We stress that what is observed (and usually fitted with power laws),
is the ``specific mass ratio'' distribution $f_{\rm obs}(q)$ which is
obtained for a sample of stars with primaries in a given mass range
($M_{p1}$ and $M_{p2}$, \citet{Kouwenhoven09}). This is certainly
the case if one studies the pairing properties of binaries whose
primary stars are white dwarfs, since the progenitors of white dwarfs
are main sequence stars with masses $1.2\lsimeq M_p\lsimeq 8 \msun$.
The observed mass ratio distribution is also usually confined to
$q\gsimeq 0.1$ because observations are often incomplete below this
value.

In what follows, we will consider the pairing possibilities listed in
section (\ref{calculations}) to see whether the existing observational
constraints, comprising the SDSS white dwarf - M dwarf wide binaries,
the local white dwarf binary sample, and the flatness of the observed
mass ratio distribution, favour any of the aforementioned routes.

\subsection{Results of calculations and comparison with observations}\label{results}

\begin{figure*}
\begin{center}
\hspace{0.1in}
\epsfxsize=0.8\linewidth
\epsfbox[26 410 581 640]{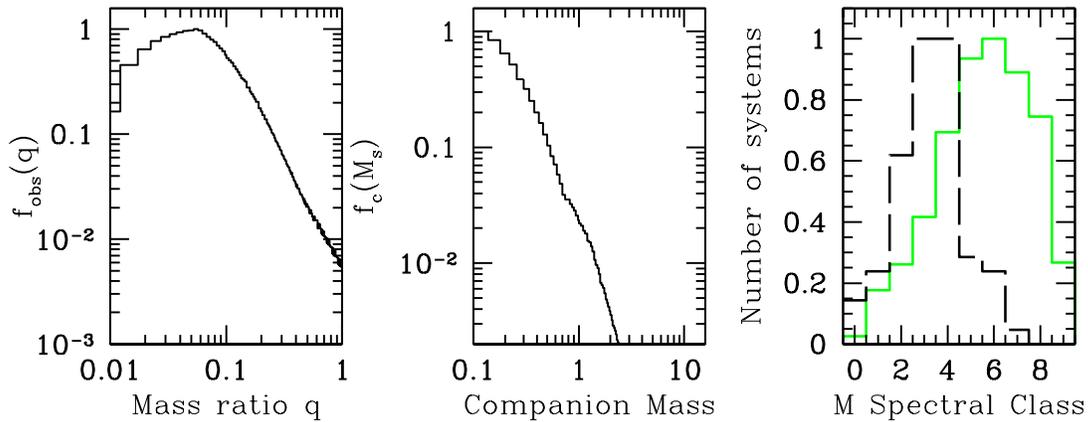}
\caption{From left to right: Mass ratio distribution, companion mass
  distribution $f_c(M_s)$ and spectral class distribution of M dwarf
  companions for a pairing where the masses of the components are
  drawn from the IMF of the type (\ref{bochanski}). The panel on the
  right shows the normalised SDSS M dwarf distribution of wide white
  dwarf-M dwarf binaries (dashed line), while the solid line shows the
  model prediction for the distribution of M dwarfs.}
\label{BS}
\end{center}
\end{figure*}

In the stellar capture scenario, single stars in dense, young clusters
form binaries via ``random pairing''. This can be simulated if we draw
the masses of both stars from an IMF distribution of the type given by
(\ref{bochanski}). The result of such pairing yields $\sim 85$ per
cent of white dwarf - M dwarf systems, in disagreement with the study
of \citet{Holberg09} which gives an upper limit of $\sim 40$ per cent
(see section (\ref{WDMS})). Such a distribution would also give a peak
in the mass ratio distribution near $q=0.2$ and a very steep decline
toward $q=1$, which is in disagreement with the observed, flat
behaviour for $0.1\lsimeq q\lsimeq 1$ discussed in section
\ref{mass_ratio_MS}. We show the results in Fig. \ref{BS}. In this
figure we also show the predicted distribution of secondary star masses
(centre panel) and M-dwarfs (right-end panel). We note that the peak
of the M dwarf distribution is near spectral class M6, while the SDSS peak
of the M dwarf distribution is near M3-M4.
We also note that if both primaries and their companions are selected
randomly from the IMF, then the outcome would be a small
percentage of Sirius-type systems ($\sim 15$ per cent) and double
white dwarfs ($\sim 2$ per cent).

\begin{figure*}
\begin{center}
\hspace{0.1in}
\epsfxsize=0.8\linewidth
\epsfbox[26 410 581 640]{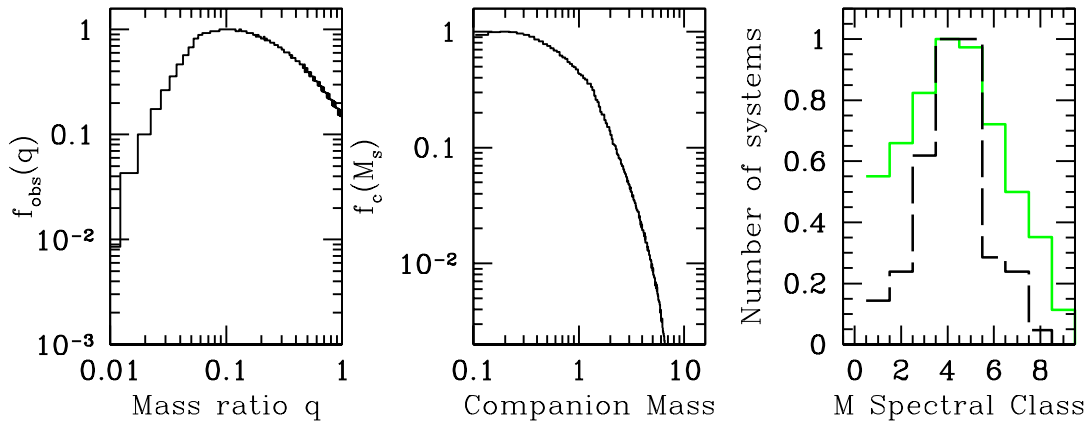}
\caption{From left to right: Mass ratio distribution, companion mass
  distribution $f_c(M_s)$ and spectral class distribution of M dwarf
  companions for a pairing of the type (\ref{gauss}) with
  $\sigma=0.4$ and $k=0.5$. The panel on the right shows the
  normalised SDSS M dwarf distribution of wide white dwarf-M dwarf
  binaries (dashed line), while the solid line shows the model
  prediction for the distribution of M dwarfs.}
\label{gaussian_fit}
\end{center}
\end{figure*}

We have also explored a distribution as given in (\ref{gauss}). Here,
the type of companion depends on the mass of the primary. The
behaviour of $f_{\rm obs}(q)$ and $M_s$ can exhibit peaks at different
locations and widths depending on the values assigned to $k$ and
$\sigma$. If we set $\sigma=0.4$ and $k=0.5$ we obtain a low
percentage ($\sim 20$ per cent) of white dwarfs - M dwarf systems and
a high percentage of Sirius-type and double white dwarf systems of 57 and
23 per cent respectively. Fig. \ref{gaussian_fit} shows the results
for this set of values. We also note that the peak of the M dwarf
distribution for the companions is predicted to occur around M3-M4, as
observed. However the slope of the mass ratio distribution for
$q\gsimeq 0.1-0.2$ is again too steep to fit the observational
evidence pointing to a flat $f_{\rm obs}(q)$.

\begin{figure*}
\begin{center}
\hspace{0.1in}
\epsfxsize=0.8\linewidth
\epsfbox[26 410 581 640]{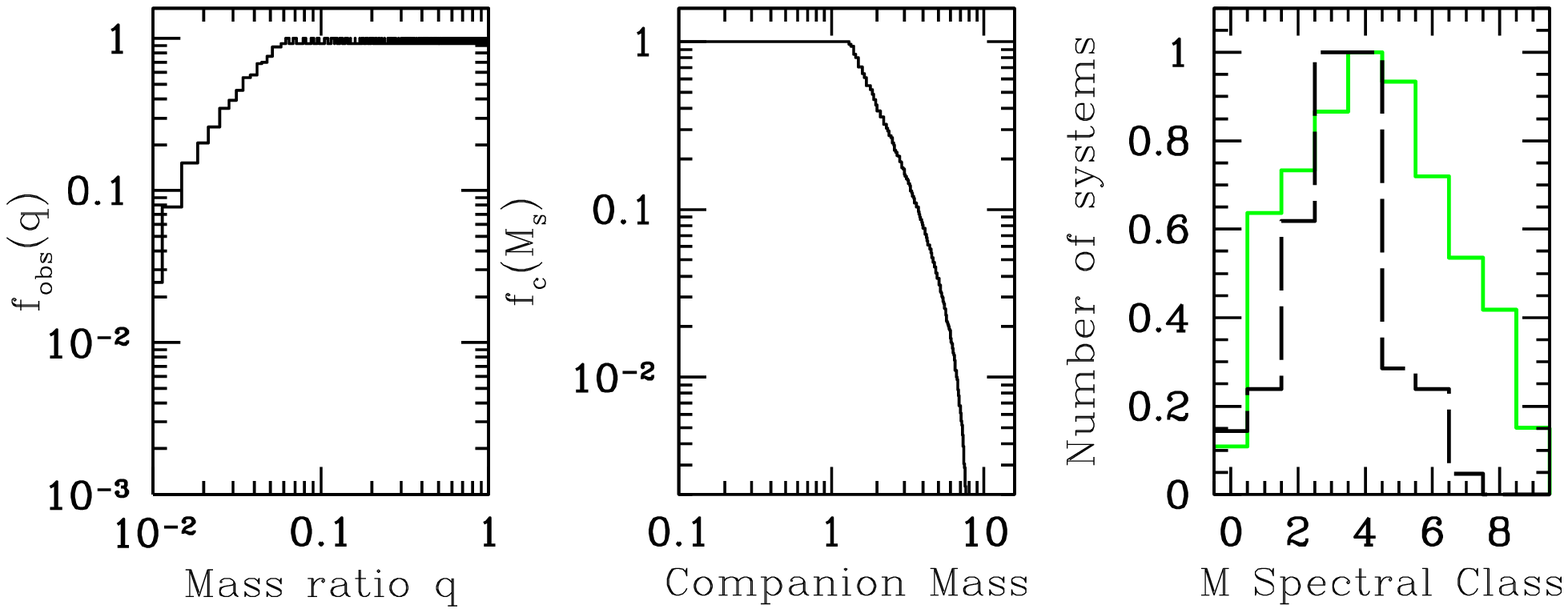}
\caption{From left to right: Mass ratio distribution, companion mass
  distribution $f_c(M_s)$ and spectral class distribution of M dwarf
  companions for a pairing of the type (\ref{qratio_1}) with
  $\beta=0$. The panel on the right shows the normalised SDSS M
  dwarf distribution of wide white dwarf-M dwarf binaries (dashed
  line), while the solid line shows the model prediction for the
  distribution of M dwarfs.}
\label{qratio_fit}
\end{center}
\end{figure*}

In the pairing cases considered above, the masses of the two stars
have been drawn from some mass distributions which yields binaries
with a certain observed mass ratio distribution. What we are going to
look at now, is the situation where the physical processes leading to
binary star formation establish the distribution of the primary mass
\emph{and} of the mass ratio distribution $f(q)$ (the ``generating
mass ratio'' distribution).  This approach on generating binaries
could be interpreted in the framework of the scale-free fragmentation
scenario for binary formation of \citet{Clarke96} who considered a
collapsing and breaking up gas cloud whose final fragments are clumps
of mass of $M_{\rm clump}$ which are distributed according to a mass
function $F_{\rm clump}(M_{\rm clump})$.  In her model, each clump
then divides into two pieces whose mass ratio distribution is
$f(q)$. That is, the fraction of clumps with mass ratio in the range
$q$ to $q+dq$ is $f(q)dq$.  In her studies, the function $f(q)$ is
independent of $M_{\rm clump}$, which is the necessary assumption for
scale-free cloud fragmentation.

In our calculations, we have considered a uniform mass ratio
distribution $f(q)$ and a power-law for the primary mass distribution
$f(M_p)$, as indicated by the SDSS data. It is important to note that
the stellar mass limits play a crucial role in the determination of
the \emph{observed} mass ratio distribution $f_{\rm obs}(q)$. In the
studies of white dwarf binaries, however, the spectral type of the
white dwarf's progenitor is restricted to be between F and late B
which gives rise to the flat $f_{\rm obs}(q)$ shown in Fig.
(\ref{qratio_fit}).  Such a pairing yields a small percentage of white
dwarfs with M dwarf companions ($\sim 18$ per cent) and a large
percentage of Sirius-type systems ($\sim 47$) and double white dwarfs
($\sim 35$). The peak of the M dwarf distribution also falls near M4,
which is in general agreement with the SDSS observations.

Finally, it may be of interest to address the fact that some
observations show the existence of a ``twin peak'', caused by stars of
similar masses pairing up.  There have been some doubts on whether
this peak is real, since binaries whose components have similar masses
tend to be brighter than those with fainter companions, thus resulting
into a possible oversampling of these systems
(e.g. \citet{Halbwachs03}). On the other hand, \citet{Tokovinin00} and
\citet{Halbwachs03} also report that the frequency of twins does seem
to be higher at short orbital periods ($< 40-50$ days). They also note
that the mass ratios of their visual binary sample with wide
separations does not exhibit a peak around $q = 1$
\citep{Halbwachs83}.

The ``twin peak'' seems also to be common among massive primaries.
\citet{Pinsonneault06} report that systems with $M_2 > 0.95M_1$
comprise 45 per cent of their population of massive binaries while the rest
exhibits a flat mass ratio distribution.

Thus, it appears that the stellar components of binaries with short
periods and some of the more massive binaries ``prefer'' similar mass
companions. This implies the coexistence of different binary formation
mechanisms.

One possibility is that in some cases, dictated by some as yet unknown
initial conditions, the scale-free fragmentation of the cloud into
clumps is then followed by vigorous disc fragmentation (see section
\ref{form_mech}). As theoretically demonstrated by the hydro-dynamical
calculations of \citet{Bate97}, such disc fragmentation would create
systems whose mass ratios approach unity. This behaviour can be
modelled by setting $k=1.0$ in equation (\ref{gauss}). In this
picture, scale-free fragmentation would then be followed by a
formation process that is dependent on the mass of the primary star.

In Table \ref{table} we summarise our results in terms of the various
systems that are generated via the pairing mechanisms considered in
this paper. This table gives a prediction of the relative percentages
of systems that one should find in a complete volume limited sample of
white dwarf binaries.

\setcounter{table}{0}
\begin{table*}
 \begin{minipage}{168mm}
 \caption{Prediction of the relative percentages of systems in a
   complete volume-limited sample of white dwarf binaries.}
 \label{tab1}
 \begin{tabular}{@{}llll}
\hline
Pairing       & Systems         &  Sirius-type systems & Double white dwarf     \\
mechanism     & with M dwarfs   &                      & systems                \\
              & (per cent)      &  (per cent)          & (per cent)             \\
\hline
Random pairing with         & 83     & 15            & 2   \\
secondaries drawn from IMF  &        &               &     \\
                            &        &               &     \\
Secondaries drawn from      & 20     & 58            & 22  \\
Gaussian distribution       &        &               &     \\
($\sigma=0.4$, $k=0.5$)     &        &               &     \\
                            &        &               &     \\
Flat generating mass ratio  & 18     & 47            & 35  \\
                            &        &               &     \\
\hline
\end{tabular}
\end{minipage}
\label{table}
\end{table*}

\section{Conclusions}

In this paper, we have seen that it is possible to gain some insights
in the behaviour of the mass ratio distribution at low values of $q$
by studying binary systems comprising of white dwarf primaries and
main sequence companions. This has allowed us, via an initial-final
mass relationship for white dwarfs, to draw some conclusions on the
pairing properties of Main Sequence of spectral type F to late B,
which are the progenitors of the currently observed white dwarfs. The
constraints that we have used are (i) the white dwarfs - M dwarf wide
binary sample from the SDSS (ii) the \citet{Holberg09} percentage of
observed white dwarf binaries in the 20 pc local sample and (iii) the
fact that the mass ratio distribution for Galactic binaries, ranging
from G to early B-type stars and even for young protostellar
associations, is either flat or slightly raising toward $q\sim
0.1-0.3$. The aim of this paper was to investigate whether it is
possible, on the basis of the data currently at hand, to explore and
possibly favour some of the proposed routes for star formation.

A prediction of our studies is the existence of a large fraction of
``hidden'' white dwarfs in binaries consisting of either double white
dwarfs or close (unresolved) Sirius-like systems.

In this context, we would also like to note that using the ROSAT Wide
Field Camera survey of the extreme ultraviolet, \citet{Burleigh98}
have revealed the existence of a previously unidentified sample of hot
white dwarfs in unresolved, detached binary systems. These stars are
invisible at optical wavelengths due to the proximity of their much
more luminous companions (spectral type K or earlier).  However, for
companions of spectral type A5 or later, the white dwarfs are easily
visible at far-ultraviolet wavelengths, and can be identified in UV
spectra. In total, 16 such systems have been discovered in this way
through ROSAT, EUVE and IUE observations. Further observations of this
kind should reveal how common these systems really are.

Clearly, also a much larger sample of white dwarf - M dwarf companions
is needed to further explore the incidence of such systems. At the
moment, the number is still too small and uncorrected for
observational biases. A clean, enlarged sample of this type is crucial
to shed more light on how binaries are formed and thus on possible
star formation mechanisms.

\section*{Acknowledgements}

The author wishes to thank Dayal Wickramasinghe for useful comments
and a careful reading of the manuscript, and Jay Farihi and John
Bochanski for their help with M dwarf distributions. The author is
also grateful to the Referee, Jay Holberg, for providing useful
information and corrections.

\end{document}